\documentclass[acmsmall,nonacm]{acmart}
\renewcommand\footnotetextcopyrightpermission[1]{} 
\usepackage[utf8]{inputenc}
\usepackage{natbib}
\usepackage{amsmath}
\usepackage{upgreek}
\usepackage{stmaryrd}
\usepackage{listings}
\usepackage{fancyvrb}
\usepackage{lstautogobble}
\usepackage{proof}
\usepackage[noabbrev,capitalize]{cleveref}
\usepackage{graphicx}
\usepackage{soul}
\usepackage{color}
\usepackage{wrapfig}
\usepackage{subcaption}

\usepackage{textcomp}

\definecolor{keywordcolor}{rgb}{0.1, 0.1, 0.4}   
\definecolor{commentcolor}{rgb}{0.4, 0.4, 0.4}   
\definecolor{symbolcolor}{rgb}{0.0, 0.1, 0.6}    
\definecolor{sortcolor}{rgb}{0.0, 0.0, 0.0}      
\definecolor{errorcolor}{rgb}{1, 0, 0}           
\definecolor{stringcolor}{rgb}{0.5, 0.3, 0.2}    

\usepackage{amsthm}

\newcommand{\x}{\times}
\newcommand{\N}{\mathbb{N}}
\newcommand{\R}{\mathbb{R}}

\renewcommand{\tt}[1]{\texttt{#1}}

\newcommand{\no}[1]{{}}

\renewcommand{\o}[1]{\operatorname{#1}}

\theoremstyle{plain}

\theoremstyle{definition}

\newcommand{\code}[1]{\lstinline[basicstyle=\small\ttfamily]{#1}}

\definecolor{a}{rgb}{0.0, .8, 0.2}    
\definecolor{b}{rgb}{0.9, 0.7, 0.1}    
\definecolor{c}{rgb}{.9, 0.3, .1}    

\AtBeginDocument{%
  \providecommand\BibTeX{{%
    \normalfont B\kern-0.5em{\scshape i\kern-0.25em b}\kern-0.8em\TeX}}}

\setcopyright{none}

\begin{document}

\title[Fast Collection Operations]{Fast Collection Operations from Indexed Stream Fusion}
\subtitle{Indexed Iterators for Everyone}

\author{Scott Kovach}
\affiliation{
  \institution{Stanford University}
  \country{USA}
}
\email{dskovach@stanford.edu}

\author{Praneeth Kolichala}
\affiliation{
  \institution{Stanford University}
  \country{USA}
}
\email{pkolich@stanford.edu}

\author{Kyle A. Miller}
\affiliation{
  \institution{University of California, Santa Cruz}
  \country{USA}
}
\email{kymiller@ucsc.edu}

\author{David Broman}
\affiliation{
  \institution{KTH Royal Institute of Technology}
  \country{Sweden and}
  \institution{Stanford University}
  \country{USA}
}
\email{dbro@kth.se}

\author{Fredrik Kjolstad}
\affiliation{
  \institution{Stanford University}
  \country{USA}
}
\email{kjolstad@stanford.edu}

\renewcommand{\shortauthors}{Kovach et al.}


\begin{abstract}
We present a system of efficient methods for traversing and combining associative collection data structures.
A distinguishing feature of the system is that, like traditional sequential iterator libraries, it
does not require specialized compiler infrastructure or staged compilation for efficiency and composability.
By using a representation based on indexed streams, the library can express complex joins over input collections while using no intermediate allocations.
We implement the library for the Lean, Morphic, and Rust programming languages and provide a mechanized proof of functional correctness in Lean.
\end{abstract}





\maketitle

\section{Introduction}

Many contemporary programming languages include an iterator interface that provides a uniform way of sequentially traversing any data type that implements it.
This approach powers the \code{for}-loop syntax available in Python, Javascript, Rust, and many other popular languages.
Some standard libraries additionally provide
combinators such as map, filter, fold, and zip to combine and transform iterators.
Compilers can leverage deforestation techniques such as stream fusion
~\citep{Wadler:1990:deforestation,GillLaunchburyPeytonJones:1993,Svenningsson:2002,coutts2007stream}
to dramatically improve the performance of such programs by eliminating redundant traversals and memory allocations.

Similar combinators can be defined for data structures that represent \emph{associative arrays}, which encode finite maps from keys to values.
However, performance trade-offs arise when using these combinators as they are traditionally defined.
We can see this with a simple example task: given a set (\code{s : Set k}) and a map (\code{m : Map k v}), find the set of keys that are present in \code{s} and associated with a value \code{v} in \code{m} that satisfies \code{v > 10}.
Consider the following API for sets and maps in the functional language Lean~\cite{moura2021lean} and three natural ways of implementing the program:

\begin{lstlisting}
    lookup    : Map k v → k → Option v
    contains  : Set k   → k → Bool
    select    : Map k v → (k → v → Bool) → Set k
    filter    : Set k   → (k → Bool) → Set k
    intersect : Set k   → Set k → Set k

    variable (s : Set k) (m : Map k v)
    -- three equivalent programs:
    def p1 := intersect s (select m (fun _ v => v > 10))
    def p2 := select m (fun k v => v > 10 && contains s k)
    def p3 := filter s (fun k => case lookup m k of
                       | None => False
                       | Just v => v > 10)
\end{lstlisting}

In the definition \code{p1}, we filter \code{m} before intersecting it with \code{s}.
This uses a potentially optimized function \code{intersect} for intersecting sets,
but comes at the cost of reifying the filtered version of \code{m} as a set and traversing it twice.
The other two programs are manually fused: they combine all operations within one traversal.
In \code{p2}, we traverse \code{m} while using \code{contains} to query \code{s},
and in \code{p3}, we traverse \code{s} while using \code{lookup} to query \code{m}.

Although \code{p2} and \code{p3} avoid the unnecessary memory allocations of \code{p1}, they have different performance depending on the relative sizes of \code{s} and \code{m}.
Identifying the superior choice may be difficult to do statically or require runtime size checks that grow more complex for larger examples.
Although \code{p1} is more modular, achieving fusion seems to require either implementations of \code{intersect} for all pairs of data structures, or restriction to a single data structure for collections.
The latter approach is often made internally within software such as database engines, but is awkward for general purpose programs.
This situation suggests that a desirable interface is missing.

In this work, we propose such an interface and implement it within three general purpose languages (Lean~\cite{moura2021lean}, Rust~\cite{matsakis2014rust}, and Morphic~\cite{brandon2023better}).
Our methods achieve fusion by internally using a data structure based on indexed streams~\cite{indexedstreams}.
Our key contribution is a fold function for indexed streams that can be ported and reliably optimized by general purpose language compilers capable of specialization.
This function elegantly ties together the essential components of an iterator library: methods to define iterators for new data types, methods for transforming and combining iterators, and methods for collecting the values of an iterator into a concrete data structure of the user's choice.
We design and implement (for each language) these library components using an interface that is only mildly more complex than traditional iterators, while enabling asymptotic speedups in computation time and greater expressiveness.



In addition to performance benefits, the small set of data concepts used in the library simplify the programming experience.
All underying data structures are wrapped in a Map-like type that we denote \code{(k →ₛ v)}.
Notationally, this emphasizes that a structure maps keys to values, just as a function does.
Sets are represented with the type \code{k →ₛ Bool},
and hierarchical data structure types are akin to curried functions: \code{k₁ →ₛ k₂ →ₛ v}.
We give several examples later to clarify and demonstrate the expressiveness of the library API.
Here, we show that \code{intersect} and \code{select} can be defined in terms of two more primitive API functions:
\begin{lstlisting}
    -- new API:
    map     : (k → a → b) → (k →ₛ a) → (k →ₛ b)
    zipWith : (a → b → c) → (k →ₛ a) → (k →ₛ b) → (k →ₛ c)

    variable (s₁ s₂ : k →ₛ Bool) (m : k →ₛ v)
    intersect s₁ s₂ := zipWith (fun x y => x && y) s₁ s₂
    select m f := map f m
    def p1 := intersect s (select m (fun _ v => v > 10))
\end{lstlisting}

Thus, \code{p1}, written exactly as before, encodes an algorithm that is fused
(the use of \code{map} and \code{zipWith} does not allocate or perform extra traversals), and that achieves worst-case optimal performance~\cite{ngo2018worst} no matter which argument is smaller.
Moreover, if its continuation is another stream operation, it avoids allocating the unnecessary \code{Set} that all three previous programs would allocate.


Our contributions:
\begin{itemize}
\item An indexed stream fold method that simplifies and generalizes previous approaches to compiling indexed streams that avoid intermediate memory allocations.
\item A design for the interfaces and methods needed to embed this in a practical library for working with associative container types (implemented in Lean, Morphic, and Rust).
\item Computer-checked proofs of correctness for the Lean library.
\end{itemize}

As evaluation, we demonstrate that the generality of the approach does not introduce a detrimental performance overhead by comparing stream-based programs to idiomatic baselines in each language.
Additionally, we give proofs of correctness for the Lean library itself, not an abstract substitute model.
Thus, if we include the Lean compiler in our trusted codebase, we have a proof of correctness for the core stream operations themselves, which gives users seeking to develop high performance, formally verified applications a strong starting point.

\section{Overview}

The paper shows how to implement a shallowly embedded library, meaning it does not require a domain-specific compiler infrastructure, for operations on associative arrays based on indexed streams. 
The library is expressive (programs syntactically resemble high level queries), has high performance (matching the runtime of hand written programs), enables seamless composition of queries and other application code (does not rely on deep embedding), and is based on rigorous principles (simple composition principles and a correctness proof in Lean).

Our approach rests on a novel formulation of the indexed stream \code{fold} function, which subsumes previous compilation algorithms.
The design is portable to multiple host languages, which we demonstrate with implementations for Lean, Rust, and Morphic.
In the following sections, we describe the approach from three points of view: the library's external user-facing interface, the internal interface for extending the library with new operators and data structures, and the interface with the host language that we rely on for good performance.
These correspond to three types of user: a programmer using one of the stream libraries, a programmer extending the library, or a programmer porting the stream library concept to a new language.

\paragraph{Outer Interface} 
In \cref{sec:library}, we define the high-level stream operators used in our library to express queries.
These operators suffice for defining relational queries built using joins, aggregates, functional predicates, and more.
We also provide traditional array programming operators such as \code{map}, \code{zipWith}, and \code{fold}.
The library is suitable for high-performance computation involving high-dimensional associative arrays, regardless of what they represent (tensors, graphs, tables, \ldots)
or what data structures are used to represent them (trees, arrays, hash maps, function closures, \ldots, or mixtures of these).

\paragraph{Inner Interface}
In \cref{sec:fusion}, we describe one way of realizing indexed streams as a data type composed of ordinary functions in Lean.
In \cref{sec:data}, we describe the suitability of the stream abstraction for several concrete data types.
Finally, in \cref{sec:fold} we discuss the design of the stream fold function and associated interfaces.
We introduce a method of constructing heterogeneous data structures for hierarchical maps out of complex stream expressions.

\paragraph{Host Language Interface}
In \cref{sec:portability}, we introduce some details from our implementations written in Lean, Rust, and Morphic.
We explain the data representation choices that enable specialization in each language.
We discuss how our high performance goal, which demands minimal copying and memory allocation, can be achieved even in Lean and Morphic, which are functional languages with pure semantics.

Until \cref{sec:portability}, all of our code examples will be taken from the Lean version of the library.

\section{Library Interface Design}
\label{sec:library}

Our library API design is motivated by the insight that many common manipulations of collection data types can be more elegantly and efficiently expressed using functions that operate on the collections as a whole.
These operations allows us to abstract away particular data structures, instead treating them uniformly as \emph{associative arrays}.
An associative array associates finitely many elements of a key type $k$ with elements of its value type $v$.
Put another way, it represents a function of type $k \to v$ that returns a default value if the key is not present, while also allowing values to be inserted and updated by key.
In the remainder, the term ``array'' will generally refer to an associative array.
In this section, we summarize the API and discuss nested collections.

\subsection{API Summary}
An iteration library needs to have three basic parts: methods to turn data structures into streams, methods to transform and combine streams, and methods to turn the stream back into a data structure.
The API described in~\cref{fig:api} presents a subset of the methods made available by our library for these purposes.
We refer to the transformation/combination methods in particular as \emph{combinators}.
The choice of combinators is inspired by prior work demonstrating that a large number of data processing tasks boil down to three categories:
\begin{itemize}
    \item applying a functional transformation to a single data collection (\code{map}, \code{filter}),
    \item joining two or more collections~(\code{zipWith}, \code{mul}), and
    \item aggregating the values of a collection into a singular value~(\code{sum}, \code{fold}, \code{eval}, \code{memo}).
\end{itemize}

These methods fully abstract away the underlying streaming representation of data.
They are sufficient to express the \emph{positive algebra}~\cite{green2007provenance}, which generalizes relational algebra to cover numeric operations and aggregates.
The small number of methods are expressive: this framework can be used to represent database queries, tensor arithmetic, dataflow analysis problems, graph queries, parsing, probabilistic optimization, and more~\cite{dolan2013fun, Elliott2019-convolution-extended}.

\begin{figure}
    \centering
    \begin{lstlisting}
        -- We use right-associative notation '→ₛ' for the Stream type.
        -- It takes two type parameters representing key and value types.
        infixr " →ₛ " => Stream
        
        -- (1) Stream construction.
        repeat  : v → (k →ₛ v)
        range   : (lo : Int) → (hi : Int) → (Int →ₛ Int)        
        -- Various data structure-specific iterators omitted; see 4.2.

        -- (2) Core combinator methods.
        map     : (f : k → a → b) → (k →ₛ a) → (k →ₛ b)
        zipWith : (f : a → b → c) → (k →ₛ a) → (k →ₛ b) → (k →ₛ c)
        filter  : (f : v → Bool) → (k →ₛ v) → (k →ₛ v)
        -- Arithmetic operators defined in terms of zipWith and fold.
        mul     : [Mul v] → (k →ₛ v) → (k →ₛ v) → (k →ₛ v)
        sum     : [Add v] → (k →ₛ v) → v

        -- (3) Core methods for accumulating the values of a stream into a collection.
        fold    : (m → k → v → m) → (k →ₛ v) → m → m
        eval    : [OfStream v v'] → v → v' → v' -- discussed in Section 4.3
        memo    : [OfStream a b] → [ToStream b c] → a → c
    \end{lstlisting}
    \vspace{-1em}
    \caption{A type-level summary of our methods for stream iteration, combination, and accumulation.
    \label{fig:api}}
    \vspace{-1em}
\end{figure}

\subsection{Nested Data}

Indexed streams are directly capable of traversing nested data efficiently.
For instance, if we have two streams of type \code{Int →ₛ String →ₛ Bool} representing sets of \code{(Int, String)} pairs, the \code{zipWith} or \code{mul} API functions can be used to efficiently traverse their intersection.
This intersection will be calculated by first intersecting the outer \code{Int} valued keys, and then for each of those calculating the intersection of the \code{String} keys.
This strategy resembles the hierarchical indices often used to accelerate database joins.
This method is well-known to guarantee worst-case optimal~\cite{ngo2014skew} performance with respect to table size.

\paragraph{Example: Matrix Multiplication}
In our Lean implementation, we provide syntactic macros that enable very concise expressions of join queries; for instance, multiplication of two sparse matrices:
\begin{lstlisting}  
    def matMul (a : I →ₛ J →ₛ v) (b : J →ₛ K →ₛ v) : I →ₛ K →ₛ v := 
      Σ j => a(i,j) * b(j,k)
\end{lstlisting}
This expression elides quite a lot of detail that the programmer need not be aware of to use the library: first, before multiplication can be done, the streams must be made the same shape using \code{repeat}; then the multiplication operator for three-level nested streams needs to be unfolded; and finally, if \code{eval} is applied to this expression, an aggregation needs to be done so that the indices $I$ and $K$ end up in the result expression, but $J$ is aggregated.
We can approximately write a fully explicit version of this function like so:
\begin{lstlisting}
    -- initially, a has shape I×J and b has shape J×K
    eval (Σ j => a(i,j) * b(j,k))
    -- each factor now has shape I×J×K
    let a' := map (map repeat) a
    let b' := repeat b
    = Σ j => a' * b'
    -- now we expand (*). we need to map the (*) function across all three stream layers:
    let mulIJK := zipWith (zipWith (zipWith (fun x y => x * y)))
    = Σ j => mulIJK a' b'
    -- finally we can eliminate the summation
    let stream := mulIJK a' b'
    = fold update (map sum (map (map (fold update) . empty) stream)) empty
\end{lstlisting}

This final line, with its baffling series of mapped operations, is calculated automatically by our implementation of \code{eval}, which extends to arbitrarily complex join expressions.
The programmer works with a simple algebraic model of relational algebra: joins and selects are product; project and aggregate are \code{sum}; data is transformed with \code{map}.
These operations are applied directly to the data structures of the program with no intermediary.
The \code{eval} function works out the minimal set of nested folds to force the expression, and the host language compiler optimizes this into a set of simple loops.
In the remainder of the paper, we will unpack the definitions needed to make this sort of computation possible.

\section{Library Internals Design}
\label{sec:internals}

Indexed streams are a simple generalization of traditional iterators,
which are structures containing one or more functions that abstract away an underlying collection to be traversed.
The sequential iterator approach is a fundamental part of many languages because iterators can express useful data transformations that ultimately compile to efficient loops with minimal memory allocation.
To implement the iterator interface for a new data structure or a new transformation method, it is necessary to understand its methods and how they are used to construct a result data structure.

In this section, our goal is to enable someone who is familiar with these issues to work with indexed stream iterators instead.
First, we explain how the data type differs from traditional iterators.
Second, we discuss two concrete instantiations of the indexed stream iterator type: one for static data structures backed by linear arrays, and one for dynamic data structures backed by trees.
Finally, we discuss a primary contribution: an efficiently compilable stream fold function that ties together iteration, transformation, and construction of nested data structures.

\subsection{Indexed Stream Data Structure}
\label{sec:fusion}

In this section, we provide background on streams and indexed streams using a series of code examples in Lean.
To motivate indexed streams, we start with traditional streams, which are more often called \emph{iterators} in modern programming languages.
There are many approaches to defining the iterator type.
In this section we start with the following one inspired by stream fusion~\cite{coutts2007stream}:
\footnote{Some Lean terminology: \code{structure} defines a new record type. An \code{inductive} is a possibly recursive algebraic data type.
The notation ``\code{(s v : Type)}'' states that \code{s} and \code{v} are types that are parameters to the \code{Step} type.
Structures can be given fields of any type,
including types (\code{s}, the stream's state type),
and functions (\code{valid} and \code{next}) whose types depend on other members.}
\begin{lstlisting}
    inductive Step (s v : Type)
      | skip (q : s)
      | emit (q : s) (value : v)
    
    structure Stream (v : Type) where
      s : Type
      q : s
      valid : s → Bool
      next  : s → Step s v
\end{lstlisting}

A stream is a state machine possessing a current state \code{q}.
When a stream is advanced by \code{next}, it produces a new state and (optionally) a value.
These two possibilities are represented in the \code{Step} type.
The function \code{valid} checks whether a state is terminal (\code{next} should not be called on such a state).
Users of the stream do not need to concern themselves with its internal state type $s$, so this is kept as a structure field rather than a type parameter.

The essential difference with indexed streams~\cite{indexedstreams} is that they are free to skip over values that are not relevant to the output of the computation.
This makes them available in computations where sequential iteration would introduce an unacceptable slowdown.
They are used to model key-value stores, so their type gains an extra type parameter \code{k}, and we make two changes to the interface to support skipping:

\begin{description}
    \item[key lower bound] When combining two or more streams, we will prefer to use ordered iteration.
    This way, a stream that is ahead in the key space can rapidly advance the others.
    To support this, we add an interface function called \code{index} that returns a lower bound on the next accessible key.
    \item[seek to key]
    To make use of these lower bound estimates, we need to replace \code{next} with a function that accepts an extra parameter that specifies a range of values that may be skipped over.
    We call the new function \code{seek}.
    It takes a key value and a boolean that indicates whether it may advance up to or past that value.
\end{description}

This is the type we arrive at:
\begin{lstlisting}
    structure Stream (k v : Type) where
      s : Type
      q : s
      valid : s → Bool
      index : s → k
      seek  : s → (k × Bool) → Step s v
\end{lstlisting}

In order to streamline later definitions and support formal verification,
we use a nearly equivalent type that corresponds to that from~\cite{indexedstreams}.
This new type prevents the stream functions from being applied to invalid states, and also eliminates the need for \code{Step}.
In Lean, given some proposition \code{P}, we can denote the type of values that satisfy \code{P} using the syntax \code{\{t // P t\}}.
We use this to constrain the domain of \code{index} and \code{seek} to only the valid states.
Additionally, we introduce a new predicate \code{ready} that delimits states that are ready to produce an output key-value pair.
This eliminates the need for the \code{skip} case of \code{step}, so we can remove it entirely.
We arrive at this final type used in our library:
\footnote{This is used in Lean, but for the Morphic and Rust implementations, we use simpler encodings that avoid dependent types}

\begin{lstlisting}
    structure Stream (k : Type) (v : Type u) where
      s : Type
      q : s
      -- predicates:
      valid : s → Bool
      ready : {q // valid q} → Bool
      -- core component functions:
      index : {q // valid q} → k
      seek  : {q // valid q} → (k × Bool) → s
      value : {q // ready q} → v
\end{lstlisting}

To produce the sequence of values in a stream, it is necessary to have a \code{next} function.
We do not include it in the interface above because it is definable in terms of the other methods: it is sufficient to use \code{next q := seek (index q) (ready q)}.
This advances the stream toward the current \code{index} if it has not yet produced a value for that key (not ready), and beyond that key value if it has.

To conclude, we note that \emph{any arbitrarily complex stream expression reduces to one instance of this data structure}.
This has the important consequence that the problem of evaluating a stream does not depend on the expression that created it.
As an example, here is a version of the stream multiplication operator:
\begin{lstlisting}
    def mul (a b : Stream k v) : Stream k v where
      s     := a.s × b.s
      q     := (a.q, b.q)
      valid := fun (qa, qb) => a.valid qa && b.valid qb
      index := fun (qa, qb) => max (a.index qa) (b.index qb)
      seek  := fun (qa, qb) i => (a.seek qa i, b.seek qb i)
      ready := fun (qa, qb) => a.ready qa && b.ready qb && (a.index qa = b.index qb)
      value := fun (qa, qb) => a.value qa * b.value qb
\end{lstlisting}

This stream computes the product of two streams by traversing the keys that they share.
The definition exhibits a pattern that is usual for stream operators:
each of the fields of the structure is defined directly in terms of those of the arguments.
This makes possible the inlining and specialization optimizations we describe later.

\subsection{Collection Data Structures}
\label{sec:data}
The library would be vacuous if the indexed stream type could not be instantiated for useful data types used in real programs.
In this section, we give an overview of its generality.
\subsubsection{Generalizing Sequential Streams}
First, we note that the indexed stream concept is at least as general as the traditional stream concept: it is always possible to take a sequential iterator for a collection data structure and wrap it with an indexed stream.
The corresponding definition of \code{seek} checks to see if the current key value is lagging behind the argument key value, and calls next if so:
\begin{lstlisting}
    seek q (i, r) :=
      let currentKey := index q
      let shouldAdvance := (currentKey < i) || (r && (currentKey = i))
      if shouldAdvance then next(q) else q    
\end{lstlisting}

However, this definition does not exploit the potential of collections with fast lookup methods.
Given knowledge of the type being traversed, we might want to implement \code{seek} so that it avoids traversing unnecessary parts~(\cref{sec:search-trees}).

\subsubsection{Sorted Linear Arrays}
Walking through a complete example is useful to review each of the indexed stream components:
\begin{lstlisting}
    def sparseLinearIterator (n : Nat) [LinearOrder k]
        (keys : Vec k n) (values : Vec v n) : k →ₛ v where
      q := (0 : Nat)
      valid q := q < n
      index q := keys[q.1]'(by simpa using q.2)
      seek q  := fun (j, r) =>
        let i := keys[q.1]'(by simpa using q.2)
        let advance := (i < j) || (r && (i = j))
        if advance then q+1 else q
      ready _ := true
      value   := fun ⟨q, _⟩ => values[q.1]'(by simpa using q.2)
\end{lstlisting}

First, the intent of this stream is to iterate through a series of key value pairs that are stored explicitly as arrays, with the first value of \code{keys} associated with the first value of \code{values} and so on.
This data structure is a near optimal representation for traversing data with a static key set.
We discuss the code line-by-line:
\begin{description}
    \item[parameters] First we have the parameters to this stream: an array of keys and an array of values.
    There are two type constraints: the arrays have the same length, \code{n}, and the key type is constrained to be linearly ordered. Besides this, the key and value types can be arbitrary.
    \item[state] The state of the stream consists of a single natural number, initially zero.
    \item[valid] The stream is valid as long as \code{q} is less than \code{n}.
    Because of this static information, there are no run time bounds checks and no possibility of out of bounds error.
    \item[index] The index function is passed a tuple: a natural number along with a proof that the number is less than \code{n}.
    The lower bound on the next possible key is simply the current key value.
    The notation \code{array[index]'proof} allows us to extract it safely by passing a bounds proof. The proof itself uses simplification to make the type of \code{q.2} match the expected type.
    \item[seek] This seek function implements linear traversal.
    It is passed a target index \code{j} and a strictness boolean \code{r}.
    It must advance if the current index is strictly before \code{j}, or if they are equal and the strictness flag is true.
    The correctness of the library does not require every seek function to advance to the given index; it only needs to make adequate progress~(see \cref{sec:correctness} for formal details).
    \item[ready] This iterator is always ready to produce its current value, so \code{ready} ignores the current state value and returns \code{true}.
    \item[value] Finally, \code{value} returns the current element of the array \code{values}.
    This function is passed both a proof of validity and a proof of readiness; since the second is uninformative, we use pattern destructuring to ignore it.
\end{description}

It is straightforward to define other array based streams, such as a sparse array that uses binary search on the key array instead of linear search (included in our evaluation), or a dense array that can be advanced in constant time.

\subsubsection{Search Trees}
\label{sec:search-trees}
A typical in-order tree traversal can be adapted in a straightforward way to implement \code{seek}.
This approach combines the benefits of iteration and key-based lookup:
because \code{seek} is passed the next key value the overall computation is interested in, it can potentially skip over subtrees with key values that are too small.
However, instead of just returning the corresponding tree value, it returns a tree node from which iteration can be resumed.

Compared to sequential iteration, this can asymptotically speed up a computation by replacing a linear factor with a logarithmic one.
Compared to approaches that traverse one data structure while repeatedly calling \code{lookup} on the other, this can also asymptotically speed up the computation by removing a logarithmic factor.
Moreover, the approach is reasonably simple, consisting of an upward pass that locates a parent node with sufficiently large key followed by a downward pass of ordinary search to find the valid successor~(see appendix for our Rust implementation).

Although we only implemented red-black trees in our three libraries, this strategy applies equally to any search-tree style data structure.

\subsubsection{Other Data Types}
We also provide methods to compute with unordered collections and implicit collections represented by functions.

\paragraph{Unordered streams}
Unordered collections can arise even when inputs are ordered.
For example, if a function is mapped over the keys of a stream, their orderedness may be lost.
Such a stream can still be traversed or aggregated.
These have limited composability: for instance, it may be inefficient to intersect two unordered collections.
We provide a \emph{memoization} operator which, among other purposes, can be used to collect the contents of an unordered stream into a data structure that supports ordered traversal.

\paragraph{Implicit streams}
It is sometimes most natural and space efficient to represent a stream as either a function or a state machine.
These are implicit in the sense of having almost no in-memory runtime representation at all.
For instance, we might express a filter predicate as a boolean function (as in the example from the introduction)
or express an indicator function such as \code{fun i j => if i = j then 1 else 0} (the identity matrix),
or access the elements of a hash-map by applying it to the keys enumerated by another stream.
We provide methods for multiplying these streams with ordinary streams to enact filters, as in the following example.

\paragraph{Example: Array slice}
Array programs typically make use of array \emph{slicing}: a way of extracting a sub-array, typically with a notation like $A[i:j]$ to denote the values of $A$ corresponding to indices from $i$ up to $j$.
For a dense array $A$, the expectation is that enumerating these values depends only on $|j - i|$, not the size of the array.

We reinterpret $[i:j]$ above as specifying an array itself that maps key values 
$\{i, i+1, \ldots, j-1, j\}$ to 1 and all other key values to 0.
Then the slice is equivalent to computing $[i:j] \cdot A$, a mask of $A$:
\begin{lstlisting}
    def slice (arr : Nat →ₛ v) (i j : Nat) : Nat →ₛ v := mul (range i j) arr
\end{lstlisting}
Despite relying on multiplication, it maintains the expected efficiency.
The cost is equal to the cost of traversing the range of indices from \code{i} to \code{j}, plus the cost of locating the first index past \code{i} inside \code{arr} and the cost of enumerating each successive value of \code{arr}.
In the case that \code{arr} is dense, these costs are constant, matching the usual behavior.
This same operation also applies to \emph{arbitrary} associative arrays, achieving predictable performance for those as well.

\subsubsection{Correctness}
As a final note to implementers of the stream interface, we provide a basic explanation of the notions of (strict) monotonicity and lawfulness~ (\cref{sec:stream-laws}).
Ensuring that your stream implementation satisfies these properties guarantees that it can be arbitrarily joined or nested with other stream types without issue.

\emph{Monotonicity} requires that the state returned by \code{seek} always has an index at least as large as the prior one.
\emph{Strict} monotonicity requires that, if \code{seek} is called at a ready state, it returns a state with strictly greater \code{index} value.
If the stream has a finite state space, this property is sufficient to guarantee termination.
Finally, \emph{lawfulness} requires that the \code{seek} function does not go too far or otherwise change the value of the stream at indices beyond the key it is passed.
See \cref{sec:correctness} for full details.

\subsection{Indexed Stream Fold}
\label{sec:fold}

Evaluating a standard iterator is straightforward: repeatedly call \code{next} and collect each value into the output data structure of choice.
Indexed streams have a similar formula: call \code{next}, and at each \code{ready} state, collect the key-value pair into the output data structure of choice.
However, indexed streams lend themselves to nested data structures, so joins across inputs with multiple levels are typical.
To reduce the stream, it is insufficient to store a linear series of key-value pairs into a data structure, because the value itself is a stream.
In this section, we describe the building blocks we use in Lean handle nested streams using only a single one-dimensional fold function.


\subsubsection{Fold}
This version of fold is taken from Lean (the definitions in Morphic and Rust are similar):
\begin{lstlisting}
@[inline] def fold (f : m → k → v → m) (s : k →ₛ v) : m → m :=
  let ⟨_, q, valid, index, seek, ready, value⟩ := s
  let rec @[specialize] go f q acc : m :=
    if valid q then
      let i := index q
      let q' := next q i
      let acc' := if ready q then f acc i (value q) else acc
      go f q' acc'
    else acc
  go f q
\end{lstlisting}

The function is quite simple, consisting of a tail-recursive loop called \code{go}.
The body of the loop checks if the stream is valid; if not, it returns the accumulator.
If it is valid and also ready, it updates the accumulator.
Whether or not it is ready, it updates the state and continues to the next iteration.

For the remaining discussion, it is sufficient to understand its type:
\begin{lstlisting}
    fold : (m → k → v → m) → (k →ₛ v) → m → m
\end{lstlisting}
This is a standard fold-like type signature: given the collection and a combining function, transform the accumulator.

\subsubsection{Data Construction}
We rely on a simple interface for incrementally constructing output collections:
\begin{lstlisting}
class Modifiable (k v m : Type) where
    update : m → k → (modify: v → v) → m
\end{lstlisting}

A collection type \code{m} is modifiable if it implements this function, which applies a function \code{modify} to the value associated with given key.
In practice, we further assume that the type \code{v} has a default value.
This way, when update is applied to a key that isn't present in a collection, this default value is passed to \code{modify}.
For nested data structures (when the collection is used to store values that are also \code{Modifiable}), this approach recovers the ``destination passing style''~\citep{shaikhha2017destination, liu, indexedstreams} well known to array language compilers.

\subsubsection{Evaluation}
We define a class called \code{OfStream} to transform streams into data collections.
This is analogous to \code{FromIterator} in Rust.
Implementations of this class record how to transform a stream into an \emph{update function} for a particular data type:

\begin{lstlisting}
  class OfStream (v v' : Type) where
        eval : v → v' → v'
\end{lstlisting}

Instead of simply giving \code{eval} the type \code{v → v'}, we give it this type because it is more natural for incrementally updating the elements of a map.
This connects more easily with the \code{Modifiable} method, as we will see shortly.
This is defined as a class so that the compiler can automatically deduce the needed instance when \code{eval} is applied to a stream/data-structure pairing.

Remarkably, only a few instances of this class are needed, with the most important being the case of a nested stream.
We write out the definition first and then discuss each of its parts:
\begin{lstlisting}
instance [OfStream v v'] [Modifiable k v' m] : OfStream (k →ₛ v) m where
    eval := fold Modifiable.update ∘ map eval
\end{lstlisting}

This method subsumes the compilation method of~\citet{indexedstreams} while being substantially more concise and portable across languages.
\footnote{There is one other instance, needed to handle aggregations, that is similarly simple but omitted for space. There is also a base case for scalar values.}
When fully unfolded, it can be compiled into a simple nest of loops.
We unpack the mechanism behind it next.

In Lean, instance parameters are written \code{[OfStream v v']}.
We read this parameter as ``values of type \code{v} can be transformed into values of type \code{v'}''.
These arguments do not need to be explicitly passed when \code{eval} is called; they are inferred by type class search.

The way to read the types in the first line of this definition is as follows:
if we can transform type \code{v} into \code{v'}, and if \code{m} is a collection type storying \code{k-v'} type pairs, then we can transform a stream of \code{k-v} pairs into a collection of type \code{m}.
When \code{v} is itself a stream type, the definition uses its \code{OfStream} instance (in the expression \code{map eval}) to recursively transform its values into the concrete values of type \code{v'} that are stored in \code{m}.
The total transformation is achieved in two steps: map \code{eval} over the values of the stream, then fold the resulting stream using \code{update}.
The first step is a stream transformation: no computation is done immediately.
The second step ultimately forces computation.

To further unpack this, using the type signatures previously given for \code{fold}, \code{update}, and \code{eval}, we can calculate the types of the subexpressions as follows:

\begin{lstlisting}
  fold        : (m → k → v → m) → (k →ₛ v) → m → m
  update      : m → k → (v → v) → m
  eval        : v → v' → v'
  map eval    : (k →ₛ v) → (k →ₛ (v' → v'))
  fold update :              (k →ₛ (v' → v')) → m → m
  -------------------------------------------------------
  fold update ∘ map eval : (k →ₛ v) → m → m
\end{lstlisting}

The apparent inefficiency of using \code{map} to create a stream of function closures is in fact never an issue: we rely on the power of the host language to perform specializing compilation to completely inline and optimize the computation.
We discuss this in greater detail in the following section.

\section{Portability Design}
\label{sec:portability}

The iterator concept has found a home in many mainstream programming languages, providing a uniform and performant interface to sequential iteration.
Similarly, indexed stream iteration can find a home in many languages and 
provide a uniform interface to \emph{non-sequential} iteration.

In this section, we explain how the indexed streams can be implemented in substantially different programming languages in such a way that their compilers ensure indexed stream fusion.
We discuss how the features of Lean, Morphic, and Rust lead to different stream type definitions that ultimately lead to similar performance and expressiveness.

As a simple example to introduce the Morphic and Rust implementations, we will use the following computation: $\sum_{i=0}^{n-1} i~(\o{mod} 5).$
Although this example does not require non-sequential iteration for optimal performance, it serves well as a minimal example for the code generation issues faced by our library.

\subsection{Lean}
As described previously, in Lean we have a single stream data type.
Each operator takes one or more streams as arguments and produces a single output stream.
Lean's dependent type system allows us to hide the stream's state type and also enforce various correctness hypotheses by construction.
These propositional arguments are erased during code generation, so they have no effect on performance.

Lean has a \code{specialize} attribute which can be used to tag a recursive function definition.
This is a request to the compiler to generate specialized versions (if possible).
We use this on a single function: the stream fold function, which is recursive and higher-order.
The remainder of the library is designed carefully so that stream constructions are always immediately available to the inline/specialize machinery.

In particular, type classes are used to resolve the \code{eval} method, which is responsible for calling \code{fold} on nested streams, described in the previous section.
Because type class resolution is performed at compile time, the Lean compiler is able to generate custom code for each invocation of \code{fold} by fully deconstructing the stream constructions passed as arguments.

Finally, we note that although Lean has pure semantics, it features an opportunistic optimization that performs mutable updates to constructors when they are held by a unique reference.
Within the Lean standard library, this feature is used to provide high performance mutable arrays and optimize list functions which would otherwise heavily allocate.
Our library cooperates with this feature: the fold and update combinators are designed to sequentially update an output data structure, so if that data structure is amenable to the mutation optimization, it will be used.

\subsection{Morphic}
Morphic~\cite{brandon2023better} is a pure functional language in the ML family.
Its compiler is capable of automatically specializing all higher-order definitions throughout a program.
In Morphic, like Lean, we use a single stream data structure.
Each stream operator defines a new element of this structure in terms of the components of the streams it is passed.
However, the stream definition in Morphic (\cref{fig:morphic-def}, left) is slightly more simple and unusual than the others.

\begin{figure}
    \centering
    \begin{minipage}{.47\textwidth}
    \begin{lstlisting}
  pub type Stream k v {
    pub Stream(
      ()        -> Bool,       // valid
      ()        -> k,          // index
      (k, Bool) -> Stream k v, // seek
      ()        -> Bool,       // ready
      ()        -> v,          // value
    )
  }

  pub sum(s : Stream k Int) : Int =
    fold(\(acc, _, v) -> acc + v, s, 0)
\end{lstlisting}
    \end{minipage}
    \begin{minipage}{.47\textwidth}
    \begin{lstlisting}
  pub range(lo: Int, hi: Int): Stream Int Int =
    Stream(
      \()     -> lo < hi,   // valid
      \()     -> lo,        // index
      \(j, r) -> range(     // seek
        max(lo, j + bool_to_nat(r)), 
        hi),
      \()     -> True,      // ready
      \()     -> lo         // value
    )
 
  pub eg(n: Int): Int =
    sum(map(\k -> mod(k,5), range(0,n)))
\end{lstlisting}
    \end{minipage}
    \caption{Stream definitions in Morphic \label{fig:morphic-def}}
\end{figure}
Notably, instead of taking in a state parameter, everything except for \code{seek} is simply a thunk which evaluates to the appropriate value.
We can motivate this design choice with the stream for the \code{range} operator used by our motivating example (\cref{fig:morphic-def}, right).
  
Note that the thunks are closed over the two parameters of \code{range}, and also that \code{seek} recursively calls \code{range} with updated \code{lo} argument.
One might expect the runtime to allocate five closures for \emph{every} iteration of this stream.
However, as we explain next, the Morphic compiler entirely eliminates this apparent cost.

In fact, the generated code that loops over the values of this stream simply allocates stack space for \code{lo} and \code{hi}, allocates no closures, and makes no calls.
When the compiler analyzes the call to fold (\cref{fig:rm-fold}, right) invoked by \code{sum} on this particular example, it is specialized to the lambda structure of \code{range}.
The structure of the closures that would otherwise be allocated is reified as an algebraic data type to hold the values that would be closed over (\code{lo} and \code{hi}).
Essentially, the call to \code{range(lo, hi)} is \emph{itself} the state of the stream.
Finally, the logic needed to extract values from this data type (corresponding to each component function) is inlined into fold, fold compiles to a loop, and the datatype itself is unboxed onto the stack.

This approach is also practically useful because the type system of Morphic does not have a feature to hide the state type.
We credit this idiom to the authors of Morphic's standard library iterator definition.

\subsection{Rust}
Rust monomorphizes all polymorphic functions during compilation.
For each instantiation of a given polymorphic function, this process generates code that is specialized to the given type arguments.
Rust also features \emph{traits}, an ad-hoc polymorphism mechanism.
A trait packages one or more types and methods together in an interface that can be implemented for multiple types.
Calls to trait methods are also monomorphized, and we use this to guarantee specialization of stream code by defining the fold function as a trait method.

Each stream-constructing operator has a corresponding \code{struct} type that holds its state.
This \code{struct} must implement the indexed stream trait.
This approach is idiomatic Rust: it mirrors the strategy of implementing \code{IntoIterator} and \code{FromIterator} from the standard library in order to provide sequential iteration of data structures and construction of a data structure after one or more iterator operations (map, filter, etc) are applied.

The running example in Rust is as follows:

\begin{lstlisting}[language=Rust]
    pub fn eg(n: i64) -> i64 {
        RangeStream::new(0, n).map(|i, _| i % 5).sum()
    }
\end{lstlisting}

It is instructive to break down the types of this expression:
\begin{lstlisting}[language=Rust]
        let s1: RangeStream = RangeStream::new(0, n);
        let s2: MappedStream<RangeStream, _, i64> = s1.map(|i, _| i % 5);
        let s3: i64 = s2.sum();
        s3
\end{lstlisting}
to note that the type of \code{s2} records the fact that \code{s2} is the result of mapping a function over a range (the underscore is a placeholder for the function's type).
It is at \code{s3} that the equivalent to \code{collect()} is called in order to realize the stream computation as a concrete value (in this case an integer).
See \cref{fig:rm-fold}, left, for the Rust version of stream fold.

Inspecting assembly output, we verified that the code generated for this expression is identical (modulo register and label names) to that of the following:
\begin{lstlisting}[language=Rust]
    let mut result: i64 = 0;
    for i in 0..n {
        result += i % 5;
    }
    result
\end{lstlisting}


\begin{figure}
    \centering
    \begin{minipage}{.47\textwidth}
        \centering
    \begin{lstlisting}[language=Rust]
    // Rust
    fn try_fold<B, F, R>(&mut self, 
                         init: B, mut f: F) 
       -> ControlFlow<R, B>
    where F: FnMut(B, Self::I, Self::V) 
             -> ControlFlow<R, B>,
    {
        let mut acc = init;
        while self.valid() {
            if self.ready() {
                let i = self.index();
                let v = self.value();
                self.next();
                acc = f(acc, i, v)?;
            } else {
                self.next();
            }
        }
        ControlFlow::Continue(acc)
    }
    \end{lstlisting}
    \end{minipage}
    \begin{minipage}{.47\textwidth}
        \centering
        \begin{lstlisting}
    // Morphic
    pub fold(f: (m,k,v) -> m, 
             s: Stream k v, acc: m): m =
        let Stream(valid, index, seek, 
                   ready, value) = s in
        if valid() {
          let r = ready() in
          let i = index() in
          let acc_ = if r { 
              f(acc, i, value()) 
            } else { acc } in
          fold(f, seek(i, r), acc_)
        } else {
          acc
        }
        \end{lstlisting}
    \end{minipage}
    \caption{The \code{fold} function in Rust (left) and Morphic (right). }
    \label{fig:rm-fold}
\end{figure}

\subsection{Discussion and Limitations}

\paragraph{Ideal Language Characteristics}
Our experience suggests that languages offering static guarantees of specialization and linear data update are most suitable to host indexed iterators.
The Lean compiler's optimizations appear to be less reliable at this time than Morphic;
Morphic's optimizer is designed around a type theoretic static analysis with completeness guarantee.
Additionally, we expect that, for example, Koka would be a reliable host language~\cite{lorenzen2023fp}, as it also implements specialization and additionally offers a type system for in-place update of pure data structures.

\paragraph{Limitations}
Our approach assumes that stream computations appear as static parts of the program text;
they cannot be generated based on runtime values or hidden behind recursive function applications.
This is a simple consequence of the static analyses for specialization that we rely on.

\section{Correctness}
\label{sec:correctness}
The Lean implementation of indexed streams allows for easy and modular correctness proofs. Prior work by \citet{indexedstreams} proposed indexed streams as an IR for a deeply embedded compiler for positive algebra contraction expressions. Although they proved that the indexed stream model faithfully computes the solution to contraction problems, an end-to-end correctness guarantee was left open. While end users could be confident of the mathematical details of the abstract indexed streams as used by their compiler, they would have to trust that the custom code generator faithfully translated indexed streams to imperative code.

However, if we directly evaluate indexed streams in Lean as in the shallow embedding of this work, the previous correctness proofs apply directly---end users now only have to trust the Lean compiler, which is widely used and well-tested, rather than a custom code generator. Moreover, end users can more easily extend and integrate indexed streams with other verified computation. In this section, we describe how we adapted the proof structure of \citet{indexedstreams} to a shallowly embedded library.

\subsection{Stream Laws}
\label{sec:stream-laws}
We briefly review the stream laws from~\citet{indexedstreams}. First, we ask that streams be monotonic, meaning that calling \code{seek} never decreases the index. We say a stream is \textit{lawful}, if, in addition to being monotonic, it satisfies the property that when the stream is at state $q$, calling seek with parameters $(i, b)$ (where $i : \iota$ and $b : \texttt{Bool}$) does not change the final \textit{semantic evaluation} (i.e. interpretation of the stream as a function from indices (keys) to values; see \ref{sec:abstract_eval}), at any index $j$ where $(j, 0) \geq (i, b)$ in the lexicographic ordering. Intuitively, calling \code{seek i} permits the stream to move up to index $i$, but it should not affect emissions whose indices are beyond $i$. In Lean, these requirements are expressed as

\begin{lstlisting}
    class IsLawful [LinearOrder ι] [AddZeroClass α] (s : Stream ι α) where
      mono : ∀ q i, s.index q ≤ s.index' (s.seek q i)
      seek_spec : ∀ q i j, (i ≤ (j, false)) → s.eval (s.seek q i) j = s.eval q j
\end{lstlisting}

If, additionally, the stream always strictly increases its index when advanced from a ready state, we say it is \textit{strictly monotonic}. Finally, we say a stream is \textit{bounded} if, roughly, for every infinite sequence of \code{seek} calls which can be made without ever terminating the stream, only finitely many of those calls are to indices past the index of the stream at the state when the \code{seek} call is made.

Most of the basic combinators we supply preserve these properties---taking bounded streams to bounded streams, monotonic streams to monotonic streams, and so on. Note that strict monotonicity can always be recovered from any bounded, disordered stream using the \code{memo} operation. This is useful if one e.g. wishes to use \code{imap} with a generic $f$.

The stream laws permit users to build sophisticated, efficient, verified computations by composing the basic combinators, while retaining confidence that these compositions of combinators remain correct. Moreover, users who wish to extend our library of combinators must only prove these properties in order to be able to automatically integrate with the rest of our library.

Finally, our shallow embedding of indexed streams and Lean's dependent type system permit us to use information about stream laws when deciding how to evaluate streams, which was not possible in previous work. For example, we could define an instance \code{OfStream \{s // s.IsStrictMono\}}, describing evaluation of the subtype of streams which are strictly monotonic, rather than simply monotonic. The evaluation function could then be optimized for the case where it is guaranteed that each index will only be seen once.

\subsection{Changes in Core Theorems}
We were able to reuse most proofs from the prior work~\cite{indexedstreams} almost directly, modulo some small changes. In particular, previously, \code{ready} produced values on all states, implicitly duplicating the work of checking if \code{valid} is true. While this was acceptable for an abstract model, we now require that a call to \code{ready} be supplied with a proof of \code{valid} to improve performance.

Similarly, other changes were made so that the indexed stream model could function as a shallow embedding of indexed streams. For example, previously, \code{valid} and \code{ready} were \code{Prop}-valued functions, not necessarily decidable. The signature of these functions was changed to return \code{Bool} to allow computation, and this necessitated a straightforward refactor of the core lemmas.

\subsection{Semantics}
\label{sec:abstract_eval}
In addition to concrete evaluation, which evaluates streams to associative arrays as described in \ref{sec:fold}, we also have a semantics which evaluates streams to finitely supported functions. If we define
\begin{lstlisting}
    def eval₀ [Zero α] (s) (q : {q // s.valid q}) : ι →₀ α :=
      if h : s.ready q then
        -- function which maps `s.index q' to `s.value q' and is 0 elsewhere
        fun₀ | s.index q => (s.value ⟨q, h⟩)
      else 0
\end{lstlisting}
then we can define the semantics as:
\begin{lstlisting}
    def eval [AddZeroClass α] (s) [IsBounded s] : s.σ → ι →₀ α
    | q => if h : s.valid q then
             -- (proof of termination elided)
             eval₀ s ⟨q, h⟩ + eval s (s.next q)
           else 0
\end{lstlisting}
This is an adaptation of the semantics given in \cref{sec:background-indexed}. Note that we have replaced the output type $\iota \to \text{Option }\alpha$ with a finitely supported function $\iota \to_0 \alpha$, taking the point of view that an associative array has a default \code{0} value which is its value at any missing key. This version is actually more general since we can always assign
$\alpha := \text{Option }\alpha'$, and set the zero value to be none

This allows us to consider streams while remaining agnostic to any particular data structure. All proofs of correctness for the core combinators in our library are given in terms of this semantic evaluation. For example, the correctness theorem for multiplication looks as follows:
\begin{lstlisting}
    theorem mul_spec (a b : Stream ι α) [IsStrictLawful a] [IsStrictLawful b]
        (q : (a.mul b).σ) :
      (a.mul b).eval q = a.eval q.1 * b.eval q.2
\end{lstlisting}

By providing a semantics which is agnostic to data structure, we allow combinators to be verified in isolation, once and for all, rather than once for every data structure that could be backing the input streams.

\subsection{Data Structure Interface}
\label{sec:data-structure-interface}

The most significant difference from prior work in the proofs is the integration of concrete data structures that model associative arrays. We need to be able to connect the semantic evaluation to the actual compilation of streams into these data structures.

In order to do this, we add a canonical evaluation from our data structure to finitely supported functions from the index type to the value type---that is, a canonical interpretation of our data structure as an associative array. Recall that our abstract semantics also provide a canonical way to evaluate a stream into a finitely supported function. Thus, the laws that \code{OfStream} and \code{ToStream} must satisfy simply assert that these evaluations commute:

\begin{lstlisting}
    class LawfulOfStream [EvalToFinsupp α γ] [EvalToFinsupp β γ] [Add γ]
        extends OfStream α β where
      eval_of_stream : ∀ (s : α) (b : β),
          (evalFinsupp (eval s b) : γ) = evalFinsupp b + evalFinsupp s

    class LawfulToStream [EvalToFinsupp α γ] [EvalToFinsupp β γ]
        extends ToStream α β where
      to_stream_eval : ∀ (d : α), (evalFinsupp (stream d) : γ) = evalFinsupp d
\end{lstlisting}

Once these laws are proved for some data structures, we can easily provide end-to-end proofs for stream computations that freely mix and match those data structures. Moreover, users can easily integrate their own data structures and verify their properties; it suffices define \code{OfStream}, \code{ToStream}, and \code{EvalToFinsupp} for the data structure and to prove that these two laws hold.

In order to make this more convenient, we have proved that for data structures implementing \code{Modifiable} satisfying some additional basic correctness properties (e.g. updating an index leaves other indices unchanged), the corresponding \code{OfStream} instance defined in \cref{sec:fold} satisfies the \code{LawfulOfStream} property. Thus, for a broad range of data structures which fit the \code{Modifiable} pattern, we get a reasonably efficient \code{LawfulOfStream} implementation almost for free.







\section{Evaluation}

To evaluate our libraries, we address two questions:
\begin{enumerate}
    \item Can the stream combinators match the performance of idiomatic, manually fused programs written in the host language, and how do their code size compare?
    \item In a version of the library with ideal performance, how much do we gain from using combinators over other traditional iteration methods?
\end{enumerate}

To address the first, we wrote three benchmarks of stream primitives and ported them to each language.
We compare the stream-based programs to one or more baseline programs written using idiomatic loop constructs.
These measure overhead when using our library to implement an identical algorithm.

To address the second, using our Rust library we (1) implemented three methods of evaluating the skewed triangle relational join query ~\cite{ngo2014skew}
and (2) compared our red-black tree stream to the red-black tree iterate/lookup strategy.
These measure algorithmic speedups that come from using fused iteration strategies over more naive strategies.

Benchmarks were compiled with \code{rustc} version 1.79, Lean4 version 4.7.0, and Morphic at commit hash d692f4cdd/llvm 16.
Tests were run single-threaded on an Intel i7-10710U;
each test measurement is an average of at least 100 runs.

\subsection{Comparison to Idiomatic Fused Loops}
\begin{figure*}[t!]
    \centering
    \begin{subfigure}[t]{0.33\textwidth}
        \centering
        \includegraphics[scale=0.28]{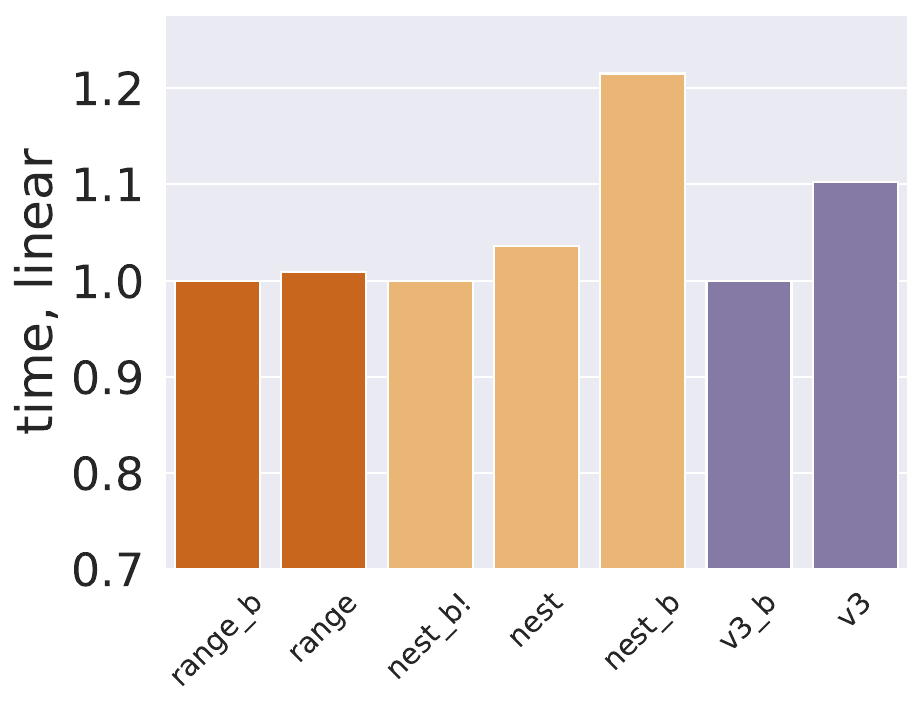}
        \caption{Rust}
        \vspace{-0.5em}
    \end{subfigure}%
    \hfill
    \begin{subfigure}[t]{0.3\textwidth}
        \centering
        \includegraphics[scale=0.28]{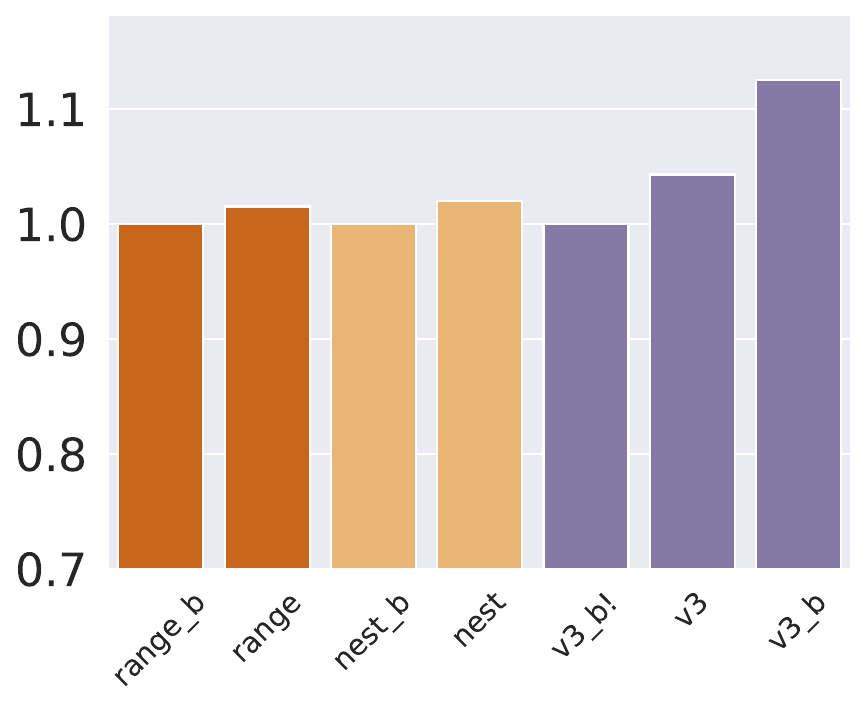}
        \caption{Morphic}
        \vspace{-0.5em}
    \end{subfigure}%
    \hfill
    \begin{subfigure}[t]{0.3\textwidth}
        \centering
        \includegraphics[scale=0.28]{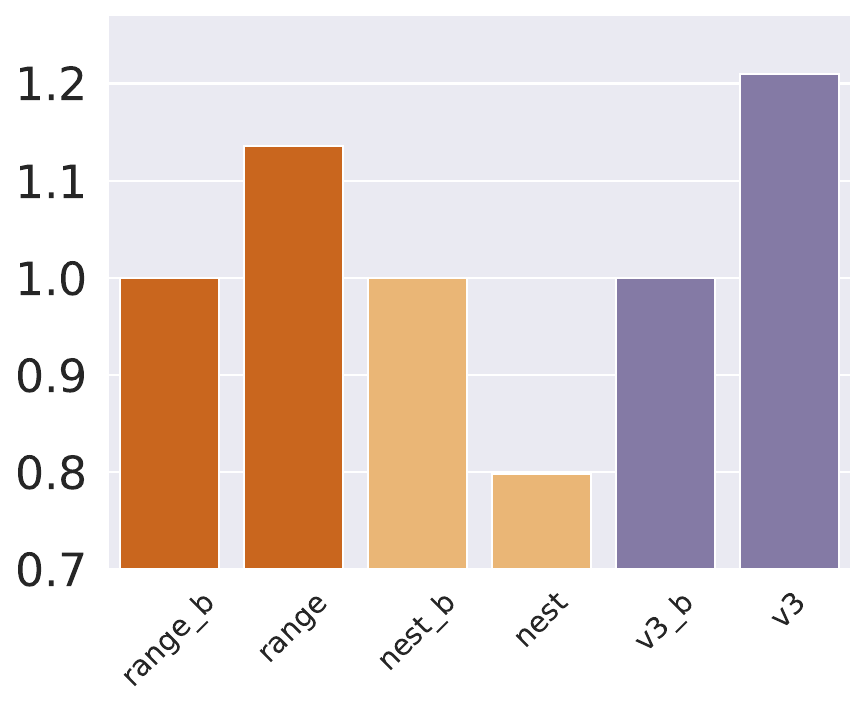}
        \caption{Lean}
        \vspace{-0.5em}
    \end{subfigure}%
    \caption{Runtimes normalized to best baseline for each test (smaller is better).
    \label{fig:perf}}
    \vspace{-0.5em}
\end{figure*}

For each implementation, we run the three tests depicted in \cref{fig:perf}.
For each of the following cases, those with suffix ``\code{_b}'' are baselines,
``\code{_b!}'' is the best baseline when more than one baseline is evaluated, and those without are computed using indexed streams:
\begin{itemize}
    \item \code{range} constructs a range iterator, maps a function over it, and sums the values
    \item \code{nest} sums the values of an $n\x n$ nested stream.
    \item \code{v3} computes the product of the values (over the intersection of keys) of three randomly generated sparse vectors. 
    We do so by simultaneously traversing all three (there is no intermediate formed).
\end{itemize}

These tests are meant to stress the library: because minimal computation is performed, we should see only the iteration overhead introduced by our generic methods.
They are reasonably complete because more complex programs are simply compositions of these basic elements: implicit stream construction, maps, and intersections.
Overall, we conclude from these tests that indexed streams introduce negligible overhead over the best baselines, and sometimes outperform simpler code that one might write in practice (see the following paragraph for discussion).

\paragraph{Implementation discussion}
The implementations using indexed streams are shorter: in every implementation, each of the three benchmarks consists of one or two expressions.
The baseline implementations range from 5--35 lines of code.
In several cases the obvious baseline turned out to be less efficient than the stream version.
For instance, it took a little trial and error to discover the second Rust program in \cref{fig:rust-sum}, which performs about 20\% faster than the first:

\begin{figure}
    \centering
    \begin{minipage}{.47\textwidth}
        \centering
    \begin{lstlisting}[language=Rust]
    // slower baseline
    pub fn nested_baseline(n: i64) -> i64 {
        let mut result: i64 = 0;
        for _ in 0..n {
            for j in 0..n {
                result += j % 5;
            }
        }
        result
    }
    
    // the stream combinator version:
    pub fn nested_iterator(n: i64) -> i64 {
        let s = RangeStream::new(0, n).map(|_, _| RangeStream::new(0, n));
        s.map(|_, v| v.map(|i, _| i % 5).sum()).sum()
    }
    \end{lstlisting}
    \end{minipage}
    \begin{minipage}{.47\textwidth}
        \begin{lstlisting}[language=Rust]
    // faster baseline
    pub fn nested_baseline_2(n: i64) -> i64 {
        let mut result: i64 = 0;
        for _ in 0..n {
            let mut a = 0;
            for j in 0..n {
                a += j % 5;
            }
            result += a;
        }
        result
    }
        \end{lstlisting}
    \end{minipage}
    \vspace{-1em}
    \caption{Rust summation baseline. \label{fig:rust-sum}}
    \end{figure}
For these cases we show both the best baseline we found and the more obvious one in the figure.

In some cases, such as the intersection of three vectors, the baseline code size scales linearly or quadratically with the number of factors involved, making that approach impractical for larger joins.
The difficulty of implementing a correct implementation also increases when different data structures are involved, especially if they have unique traversal methods.
When data types change, the stream programs remain the same: the \emph{only} difference is in input and output type annotations.
Finally, we reiterate that the stream functions compose, whereas the baseline approach requires writing a manually fused program for each new problem.

The programs in Rust and Morphic are somewhat more verbose than those in Lean, because we did not implement an equivalent to the algebraic notation and \code{eval} present in the Lean library.
For Rust, we believe this would be possible using additional traits and Rust macros.

\paragraph{Variance across languages}
In our tests, Rust performed best in absolute terms, with about a 2x edge over Morphic.
Lean was less consistent: seemingly small changes can easily affect inlining behavior and made benchmarking relatively challenging.
We also struggled to consistently ensure application of the constructor reuse optimization, which is needed to ensure that no redundant memory allocations occur.

\subsection{Comparison to Traditional Iterators}
For our second test, we use the Rust implementation to implement two problems: the triangle join query and a test of red-black tree traversal.

\paragraph{Triangle Join}
The triangle join query is the following relational algebra expression: $R(a,b)\bowtie S(b,c)\bowtie T(a,c)$.
This benchmark tests many components of the library: nested iteration of sparse data structures, construction of a nested output data structure, and sparse intersections.
We use it to illustrate that order-of-magnitude gains can be made when passing from a naive implementation to an unfused stream-based implementation and then to a fused stream implementation.
The data for this benchmark is stored using sparse arrays as a variant of CSR format.
We use three relations each containing 20,000 pairs of Strings.

\begin{figure}
    \centering
    \begin{subfigure}[t]{0.45\textwidth}
        \centering
        \includegraphics[scale=0.35]{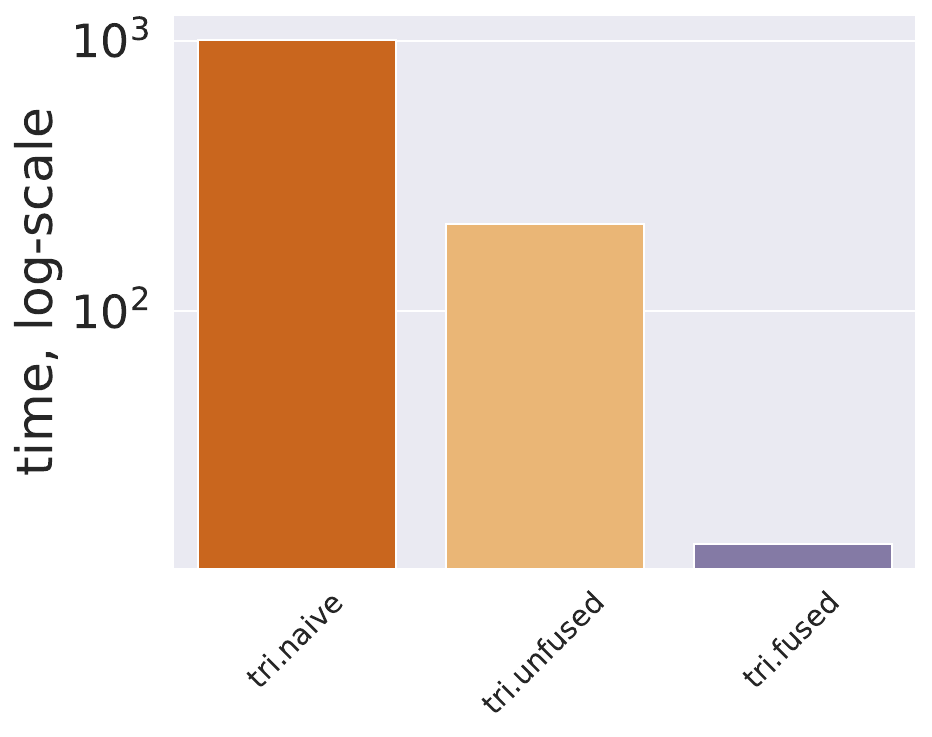}
        \caption{Triangle Query \label{fig:rust2}}
        \vspace{-1em}
    \end{subfigure}%
    ~
    \begin{subfigure}[t]{0.45\textwidth}
        \centering
        \includegraphics[scale=0.35]{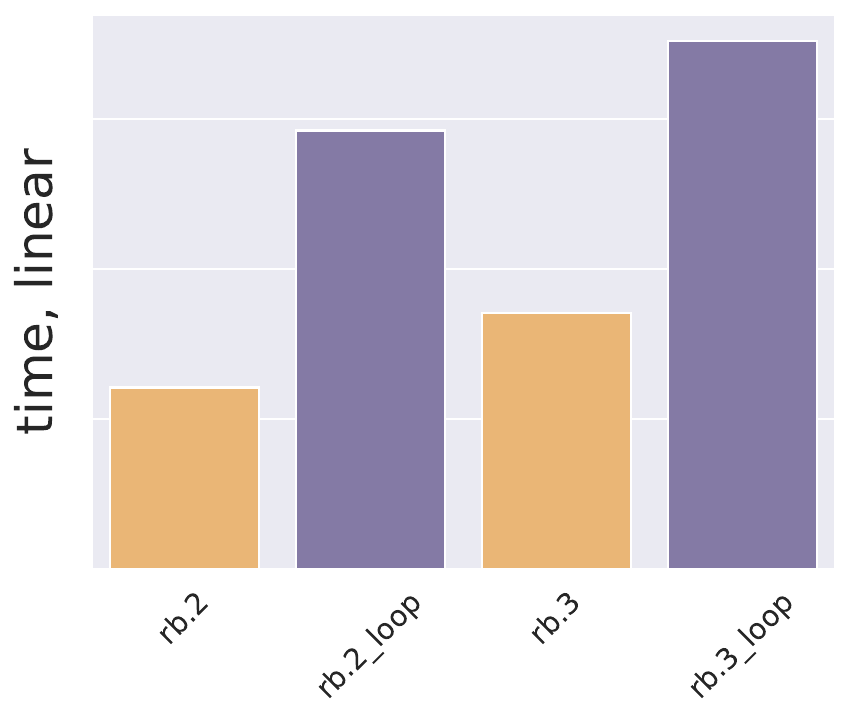}
        \caption{Red-black Tree \label{fig:rust3}}
        \vspace{-1em}
    \end{subfigure}%
    ~
\end{figure}

The \code{tri.naive} implementation measured in \cref{fig:rust2} is an implementation of iterate and lookup: it uses standard Rust iterators to get values of $a$ and $b$ from $R$, 
then does a lookup on $b$ into $S$, traverses the corresponding values of $c$, and finally uses lookup to check if $a,c$ is an element of $T$.
The relatively high cost of this method arises from the many binary searches done for the lookups.

The second implementation (\code{tri.unfused}) performs pairwise joins using galloping binary search: we use indexed streams to first join $R$ and $S$ into a temporary relation, and then join this temporary with $T$.
The speedup of 5x demonstrates the value of incrementally scanning and initiating fewer searches from scratch.

The final implementation (\code{tri.fused}) is fused: it computes the $a$ attribute values, then the $b$ values, and then the $c$ values.
This method also uses streams that perform galloping binary search, but it only traverses each input a single time and avoids allocating an intermediate result that is asymptotically larger than the final result value, leading to a speedup of about 20x in this case.
This is the default approach that results from using stream combinators to tackle this problem.

\paragraph{Joint Tree Traversal}
When intersecting the keys of two or more trees, we can either traverse one and access the other by key, or simultaneously traverse them as indexed streams.
The relative performance is shown in \cref{fig:rust3}, where \code{rb.2} (intersection of two trees) and \code{rb.3} (intersection of three trees) are the streaming tests.
These intersect red-black trees containing one million random integer keys each.
The streaming approach is faster by about 2x.

\section{Related Work}

\begin{wrapfigure}{r}{0.3\linewidth}
    \centering
    \vspace{-2.5em}
    \includegraphics[width=\linewidth]{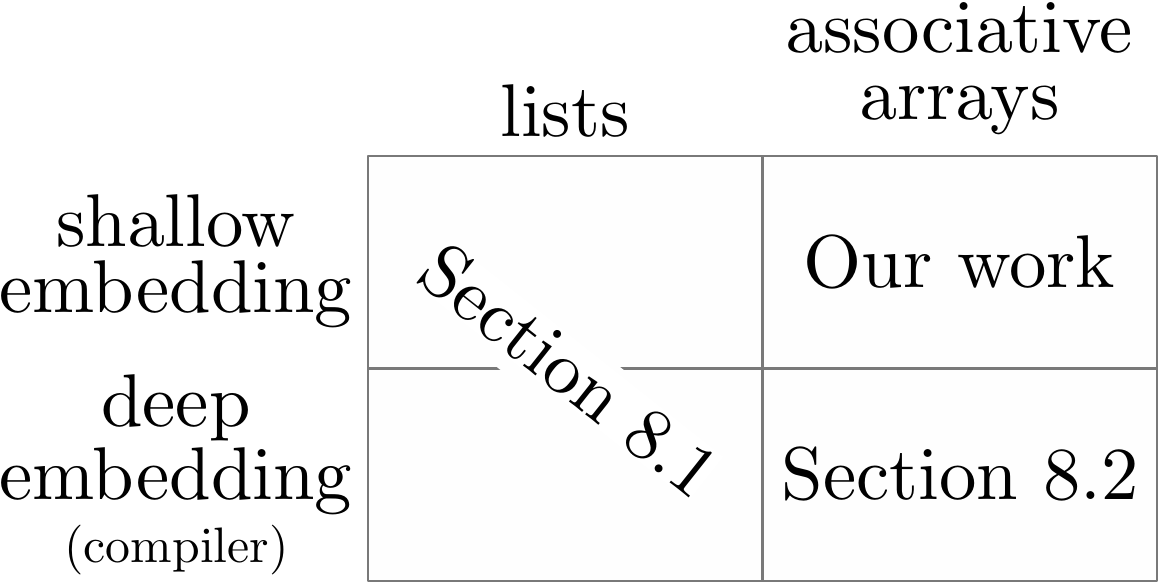}
    \vspace{-2em}
    \caption{
        \vspace{-1em}
        Related work overview
        \label{fig:related-work}
    }
\end{wrapfigure}

Indexed streams generalize the idea of stream fusion, which applies to operations on lists, to operations on associative arrays. We discuss prior work on stream fusion, as well as work on deeply embedded libraries for operations on dictionaries and their compilers. \cref{fig:related-work} gives an overview of how our work relates to prior work on shallowly embedded stream fusion, and prior work on deeply embedded approaches to domain-specific compilation of operations on associative arrays.

\subsection{Stream Fusion on Lists}

The idea of fusing combinators dates back to Wadler’s early work on deforestion~\cite{Wadler:1990:deforestation}, where combinators such as \code{sum}, \code{map}, and \code{upto} could be fused if expressed in a treeless form. By contrast, the work by \citet{GillLaunchburyPeytonJones:1993} allows any programs as input and uses two combinators \code{build} and \code{foldr} for fusion, enabling many standard list functions. The \code{build}/\code{foldr} rule cannot, however, handle a \code{zip} combinator or functions that use accumulators, such as \code{foldl}. \citet{Svenningsson:2002} takes a different approach, using \code{unfoldr} (to build lists) and \code{destroy} (to consume lists), which enables the use of accumulator functions but cannot handle nested lists or list comprehension.

\citet{coutts2007stream}  propose a generalized stream approach that solves the above problems, including handling of \code{zip}, \code{concat}, and list comprehension. In contrast to previous work, this approach uses an explicit co-structure: the \emph{Stream} type. The key idea is to use two functions, \emph{stream} and \emph{unstream}, to convert between lists and streams and eliminate such pairs using rewriting rules. Another idea introduced in this work is to use Skip, which is not necessary for the semantics but essential for the implementation. The stream fusion work is implemented as a Haskell library, making use of rewrite rules for eliminating  \emph{stream}--\emph{unstream} pairs. Similar to our work, stream fusion relies heavily on standard optimization phases in the compiler to eliminate temporary data.

There are several extensions to the stream fusion approach presented by \citet{coutts2007stream}. Generalized stream fusion~\citep{mainland2013} extends the ideas of stream fusion with a way to bundle multiple stream representations, enabling vector instructions and SIMD computation. Although the  \code{build}/\code{foldr} approach by \citet{GillLaunchburyPeytonJones:1993} supported the \code{concatMap} function, the original stream fusion work was implemented as user-level rewrites in GHC, and could not handle \code{concatMap}. \citet{10.1145/2543728.2543736} presents a solution that turns \code{concatMap} into \code{flatten}, a combinator that is specialized on a fixed state type and step function for the inner stream. Moreover, rewrites are performed using the GHC plugin HERMIT. In our setting, programs can instead be written naturally directly with a flattened style.
\citet{kiselyov2017stream} present a library approach called \emph{strymonas} that enables both guaranteed high performance and expressiveness. 
Their approach is based on staging, in this case, metaprogramming using LMS~\cite{RompfOdersky:2010} for Scala and BER MetaOCaml~\cite{kiselyov2014design}. Similar to the above approaches, our representation can encode several combinators for stream fusion, including, e.g., \code{zip} and \code{map}.
Instead of explicit staging, we use specialization, enabling a limited form of partial evaluation. 
In contrast to the work on stream fusion, our work applies to associative arrays.


\subsection{Fusion of Associative Arrays using Deep Embedding}


Over the last ten years, several research groups have explored compiler support for deeply embedded sparse collection languages~\cite{indexedstreams,shaikhha2022functional,kjolstad2017tensor,ahrens2023looplets}. In these systems, operations on sparse collections such as tensors, dictionaries, and relations, are implemented separately from the host language, for example as a LLVM-based runtime compiler inside a C++ library.

Our indexed stream fusion approach builds on the ideas of the indexed streams of \citet{indexedstreams}. Their Etch system, however, is a deeply embedded library that compiles contraction programs written in a separate domain-specific language. It also compiles contraction programs directly to C code, and uses the indexed stream abstraction only for verification. In contrast, we show how indexed streams can be written as an extensible library and how fusion can be carried out by the host language itself.

\citet{shaikhha2022functional} proposed a functional language (SDQL) over semi-ring dictionaries and showed how it expresses both relational and linear algebra. They also propose a deeply embedded compilation approach that fuses semi-ring operations over these dictionaries. Unlike indexed streams, SDQL is restricted to collections that support lookup (e.g., a hash map). \citet{Shaikhha:2022} also showed how to fuse recursive operations on monoid dictionaries, which can represent sparse linear algebra (e.g., the nested operations required for matrix multiplication). The fusion approach removes temporaries within the recursive operations, relies on staged compilation, and restricts the underlying dictionary data structures.

Another line of work on compilers for sparse collections are on compiling sparse linear/tensor algebra. The first approach was the MT1 compiler of \citet{bik1993compilation,bik1993advanced,bik1996automatic}. The TACO compiler~\cite{kjolstad2017tensor} developed a different approach that generalized to sparse tensor algebra and supported fusion, by generating co-iteration code over sparse data structures. It also supported user-defined data structures~\cite{chou2018} and scheduling~\cite{senanayake2020}. The Finch compiler~\cite{ahrens2023looplets} further generalized the co-iteration approach. These systems, however, are limited to sparse tensor algebra and are also implemented as deeply embedded compilers inside C++ or Julia libraries, in large part due to their complexity.

Finally, the C\# language provides the LINQ library~\cite{linq} for querying in-memory collections and other iterables. The LINQ query language is SQL, which is higher level than our collection operators and can be built on top of them. Furthermore, users can insert new data structures into LINQ by overloading the Enumerable interface, but the execution strategy is left to client code, so any domain-specific compilation approach must be implemented as a deeply embedded compiler.

\section{Conclusion}
In this paper we have shown how to develop a library that makes indexed iteration over associative data structures as convenient as sequential iteration without sacrificing performance.
We hope that this technique might become as pervasive as the iterator
so that users are free to write correct, performant data processing applications that avoid the friction of piping data into a database instance or domain specific language.
We believe this work puts associative array manipulations on a solid foundation, enabling 
library development for general purpose languages so that specialized compilers are no longer needed.

\bibliographystyle{ACM-Reference-Format}
\bibliography{sample-base}


\appendix

\section{RBTree seek in Rust}
\begin{lstlisting}[language=Rust]
    /// Continue advancing like next() but skip over nodes designated by cmp_fn
    /// cmp_fn should be downwards closed i.e., if cmp_fn(a), then for all b < a, cmp_fn(b)
    /// Takes O(log n) time, where n=size of tree
    #[inline]
    fn seek(self, cmp_fn: impl Fn(&K) -> bool) -> NodePtr<K, V> {
        if self.is_null() || !cmp_fn(unsafe { &(*self.0).key }) {
            return self; // minor edge case: never advance backwards
        }

        // let `target` be the least node not satisfying `cmp_fn` (the one we want to return)
        // Let `ancestor` be the least ancestor of `candidate` greater than it.
        // We maintain the following loop invariants:
        // 1. If target < ancestor, either target = candidate or target is a descendant of current
        // 2. current < ancestor
        // Note that these invariants always hold if current = candidate.right() and target > candidate
        let mut candidate = self;
        let mut current = self.right();
        // In principle, this loop should be `while !candidate.is_null() && cmp_fn(unsafe { &(*candidate.0).key })` (i.e., target > candidate).
        // However, for efficiency, we instead write this as a loop with explicit breaks when the condition is false
        // (We've also explicitly checked the condition at the beginning of the function)
        loop {
            // For optimization, tell the compiler that the loop condition holds explicitly
            if candidate.is_null() { unsafe { unreachable_unchecked() }; }
            let parent = candidate.parent();
            let is_left_child = parent.is_null() || parent.left() == candidate;
            // We will assign candidate to its parent, but in the comments below, `candidate` refers to the old value of candidate.
            // From target > candidate (loop condition), the loop invariant becomes: if target < ancestor, target is a descendant of current
            // (target = candidate is impossible)
            candidate = parent;
            if is_left_child {
                // Case 1: candidate is a left child (so ancestor = parent)
                if candidate.is_null() || !cmp_fn(unsafe { &(*candidate.0).key }) {
                    // Case 1a: target <= parent
                    // The loop invariant says if target < parent, target is a descendant of current.
                    // Thus, either target = parent or target is a descendant of current.
                    // Also, (2) clearly continues to hold since ancestor can only increase.
                    // Thus, the loop invariants continue to hold with candidate := parent (and the loop ends).
                    break;
                } else {
                    // Case 1b: target > parent
                    // In this case, assign current = parent.right() and continue the loop.
                    // The invariants hold because target > parent (just like in initialization).
                    current = candidate.right();
                }
            } else {
                // Case 2: candidate is a right child
                // In this case, ancestor does not change when candidate := parent since parent < candidate
                // so the invariant continues to hold with candidate := parent. Similarly, we don't
                // need to check the loop condition to see if we should break since target > candidate > parent.
                // This entire case could be a no-op. For optimization purposes, however, we do one step
                // of a binary search for target from current
                // (which maintains the invariant that if target was a descendant of current, it still will be).
                if !current.is_null() {
                    if cmp_fn(unsafe { &(*current.0).key }) {
                        current = current.right();
                    } else {
                        // If we find target <= current, then since current < ancestor
                        // we know target is a descendant of current.
                        // Thus we can end the loop early (you can check that the invariants still hold).
                        candidate = current;
                        current = current.left();
                        break;
                    }
                }
            }
        }

        // At the end of the loop, we must have target <= candidate. Thus, target < ancestor, 
        // so either target = candidate or target is a descendant of current. We do a binary search from current for target.
        while !current.is_null() {
            if cmp_fn(unsafe { &(*current.0).key }) {
                current = current.right();
            } else {
                candidate = current;
                current = current.left();
            }
        }
        candidate
    }

\end{lstlisting}
\section{Background: Stream Fusion to Indexed Streams}
\label{sec:background}

This section contains background material on stream fusion~\cite{coutts2007stream} and indexed streams~\cite{indexedstreams}.
It is a more gradual version of \cref{sec:fusion}.
We explain their relationship by starting with streams representing sequences of data and gradually augmenting their definition until we arrive at a representation for streams that represent associative arrays.
Thus, we show how indexed streams can be conceived of as a gradual evolution from sequential streams.
We will highlight a few points where further attention is required to enable application of the indexed stream concept to practical functional programming.
All code examples in this section are in Lean.

\subsection*{Sequential Stream Fusion}

Stream fusion~\cite{coutts2007stream} is an approach to fusing computations on sequential data structures such as lists and strict arrays.
Its key idea is to replace the recursive list definition
with an iteration function that maps a state to an optional value and successive state.
The iteration function must be able to return a special value to signal termination.
The following code defines an inductive datatype \code{Step} with three constructors\footnote{The constructor $\texttt{emit}$ has the same meaning as $\texttt{yield}$ in the original paper on stream fusion~\cite{coutts2007stream}. We use the name $\texttt{emit}$ to make it consistent throughout the paper.
}
and a structure (record) \code{Stream} with named fields:

\begin{lstlisting}
inductive Step (σ α : Type)
  | done
  | skip (state : σ)
  | emit (state : σ) (value : α)

structure Stream (α : Type) where
  σ : Type
  q : σ
  next : σ → Step σ α
\end{lstlisting}
A \code{Stream}  contains a state type $\sigma$, a state \code{q}, and transition function \code{next}.
At a given state, the \code{next} function may return \code{done},
signaling the absence of any more values in the stream,
\code{skip}, indicating that no value is produced at the current state and yielding a new state,
or \code{emit}, yielding a value and successor state.

Now a function such as \texttt{map} can be defined non-recursively:

\begin{lstlisting}
def map : (α → β) → Stream α → Stream β := fun f s => {
  s with
  next := fun state =>
    match s.next state with
    | done             => done
    | skip state       => skip state
    | emit state value => emit state (f value)
}
\end{lstlisting}
The function \code{map} returns a structure of type \code{Stream} by updating field \code{next} of \code{s} using the \code{with} construct.
This transformation to non-recursive form aids the optimizer: it is more straightforward to inline successive operations, potentially avoiding both list cell allocations and allocations from user code.
The pioneering stream fusion work~\cite{coutts2007stream} showed that standard compilation techniques are sufficient to transform pipelines built from map, filter, zip, and other operations into a single loop.

The streams we have seen so far can be given a semantics in terms of lists:

\begin{lstlisting}
partial def eval : Stream α → List α := fun {q, next} =>
  match next q with
  | done => []
  | skip q => eval {q, next}
  | emit q value => value :: eval {q, next}
\end{lstlisting}
This implements the correspondence between \code{done} with the empty list and \code{emit} with list cons. The example function is \code{partial}, to allow general recursion without requiring Lean to prove termination.

\subsection*{Sequential Container Fusion}
\label{sec:background-indexed}

After evaluating a stream, each value is equipped with an implicit position within the overall output sequence;
this position is of semantic relevance for combinators such as \code{zip},
which pairs together the output values of two streams one by one. This positional semantics works fine for sequences, but we are building towards a stream representation for data structures encoding associative arrays.
In this setting, each output value is associated with a particular key, or \emph{index}, not its position.

Taking inspiration from \citet{indexedstreams}, we make a new definition with a type parameter $\iota$, values of which we call \emph{indices},
and we allow the new \code{emit} constructor to produce both an index and a value:
\begin{lstlisting}
inductive Step (σ ι α : Type) where
  | done
  | skip (state : σ)
  | emit (state : σ) (index : ι) (value : α)
\end{lstlisting}
This change allows for streams that do not produce output at every index.
Semantically, we want a stream using this new \code{Step} type to represent an associative array,
which is most easily specified abstractly as a mapping of type $\iota \to \textrm{Option}~\alpha$.
We define the abstract semantics like so:

\begin{lstlisting}
-- a map with no entries
def zero : ι → Option α := fun _ => none

-- a map with one entry
def singleton : ι → α → ι → Option α := fun i v x =>
  if x = i then some v else none

-- a map containing all entries yielded by the stream
partial def eval [Add α] : Stream ι α → (ι → Option α) := fun {q, next} =>
    match next q with
    | done => zero
    | skip q => eval {q, next}
    | emit q i v => (singleton i v) + eval {q, next}
\end{lstlisting}
Notice the close similarity to the previous \code{eval} function.
The result of \code{eval} is a function with finitely many non-\code{none} values.
We require that the values can be combined using \code{Add} in case the stream yields the same index at multiple steps.
In contrast to the previous list-based \code{eval} function, this one does \emph{not} produce an efficient representation of its output (the result is a sum of closures).
Nevertheless, it mirrors the structure of the efficient one we will derive in a later section.

This representation is already sufficient to fuse \tt{filter} and \tt{map} operations.
These operations operate on \emph{all} the elements of \emph{one} collection,
so there is no harm in treating it as a sequence to apply them.

\paragraph{Intersection/Multiplication}
Now we turn to combining data from multiple streams.
The most widely used data processing combinator is probably the \emph{natural join}.
Prior work has shown that join can be obtained in full generality from a simple building block: the product of two streams, which yields values at indices shared by both streams.
That is, when the value type $\alpha$ has a multiplication operator satisfying $0\cdot a = a\cdot 0 = 0$, there is a natural way of lifting it to act on an entire stream.
A key insight of our work is that this single operation generalizes surprisingly well, enabling both efficient compilation and new operations and applications, including relational join and sparse tensor operations.
Here is our first pass at a definition:

\begin{lstlisting}
def mul [Mul α] : Stream ι α → Stream ι α → Stream ι α := fun a b => {
  q := (a.q, b.q),
  next := fun q =>
    match a.next q.fst, b.next q.snd with
    | done, _                     => done
    | _, done                     => done
    | skip q₁, _                  => skip (q₁, q.snd)
    | _, skip q₂                  => skip (q.fst, q₂)
    | emit q₁ i₁ v₁,
      emit q₂ i₂ v₂ =>
      match compare i₁ i₂ with
      | .lt                       => skip (q₁, q.snd)
      | .gt                       => skip (q.fst, q₂)
      | .eq                       => emit (q₁, q₂) i₁ (v₁ * v₂)
}
\end{lstlisting}
The correctness of this function relies on \emph{strict monotonicity} of the input streams,
the assumption that the sequence of indices yielded by \code{emit} always increases.
The function merges together the index sequences for \code{a} and \code{b},
emitting the product of values whenever indices match.
Notice that when either input is sparse (non-zero at relatively few values of $\iota$)
the output is likewise sparse, because it only yields values at indices in the intersection.

\paragraph{A small refactor}
\label{par:small-refactor}
However, this definition computes the intersection inefficiently.
The issue is that it inevitably traverses \emph{all} of the states of at least one stream before terminating.
In the final case branch, we know that the \emph{larger} of $i_1$ and $i_2$ is a lower bound for the next non-zero output.
If the other stream is traversing a realistic data structure, it likely supports a way of quickly finding that index.
Thus, there is a missed opportunity to quickly advance or \emph{seek} the stream forward.

Before describing a solution, we will make a semantics-preserving change to the stream definition.
The goal of this change is to remove the need for the \code{Step} type,
which ultimately simplifies the definition and compilation of combinators down the line. We call this a \textit{sequential stream}:

\begin{lstlisting}
structure Stream (ι α : Type) where
  (σ : Type) (q : σ)
  valid : σ → Bool                   -- "done"
  next  : {x : σ // valid x} → σ
  ready : {x : σ // valid x} → Bool  -- "skip"
  index : {x : _ // ready x} → ι      -- "emit"
  value : {x : _ // ready x} → α     -- "emit"
\end{lstlisting}

We have labeled each function with the \code{Step} constructor it approximately corresponds to (\code{next} still corresponds to itself).
In addition to explaining why this is equivalent to the previous \code{Stream} definition,
we take this opportunity to explain some Lean notation.
In particular, the notation \code{\{x : T // P x\}} is notation for a \emph{subtype}:
an element of \code{\{x : T // P x\}} is a pair of a value $x : T$ and a proof that $P$ holds for $x$.
These proofs are eliminated during code generation.
The type $T$ can be inferred from the type of $P$, so it is generally possible to omit it.

The \code{valid} predicate is used to signal termination.
When the \code{next} function of the previous formulation would return \code{done},
in the present formulation it returns a state $q'$ where \code{valid q' = false}.
The \code{ready} predicate, which can be applied only to valid states, marks states that yield output.
When the previous stream would return \code{skip q'}, instead here we have \code{ready q' = false}.

Finally, the successor state and yielded index/value pair are accessed using the remaining three methods.
In the prior case of \code{next q = emit q' i v}, we have
\code{next q = q'}, \code{index q' = i}, and \code{value q' = v}.

\subsection*{Indexed Streams}
\label{sec:thepoint}
We can present the final indexed stream formulation:
\footnote{In order to avoid confusion with the stream fusion \lstinline[basicstyle=\footnotesize\ttfamily]{skip} constructor, we use the term ``seek'' for what is called ``skip'' in prior work~\cite{indexedstreams}.}

\begin{lstlisting}
structure Stream (ι α : Type) where
  (σ : Type) (q : σ)
  valid : σ → Bool
  index : {x // valid x} → ι              -- available whenever valid
  seek  : {x // valid x} → ι × Bool → σ  -- replaces `next`:
  ready : {x // valid x} → Bool
  value : {x // ready x} → α
\end{lstlisting}

Notice that we have only made two small changes:
\begin{enumerate}
\item addition of an index and boolean parameter to \code{next},
which we rename \code{seek}
\item \code{index} returns a value at all valid states, not just ready states.
\end{enumerate}

As previously noted, the purpose of the $\iota$ parameter to \code{seek} is to allow \code{seek q (i, b)} to return a state that makes progress toward a state $q'$ with \code{index} at least $i$; it may skip any states with lesser index.
The boolean parameter essentially indicates whether the stream may advance past the given index (this technical detail plays an important role during evaluation and proving correctness, which we elaborate later on).

The latter change to \code{index} is simple: we require that \code{index} provides a lower bound on the next non-zero value.
When two data structures are traversed simultaneously, as in the multiplication combinator, this index information can enable the \code{seek} method of the secondary stream to advance more effectively, even when neither stream is ready to provide a value.

These two changes in fact generalize the previous definition. Given \code{seek}, the canonical definition of \code{next} is \code{next q := seek q (index q, ready q)}, which asks the stream to make progress towards moving past the index it is currently at. Conversely, given \code{next}, we can always define \code{seek} by simply calling \code{next} if the requested index is at least the current index (and doing nothing otherwise). Many data structures, however, offer more efficient implementations of \code{seek}.


\paragraph{Hierarchical data}
Finally, since a stream encodes a particular associative array (which can be thought of as a function), we encourage the reader to regard a stream as a sort of function.
We adopt the notation $\iota \to_s \alpha$ for the type $\textrm{Stream}~\iota~\alpha.$
This suggests the possibility of \emph{nested stream types} such as $\N \to_s\N\to_s\R$; this is a stream whose values are streams whose values are real numbers.
Such streams can encode hierarchical data, for instance matrices, graphs, or data tables.

Array programming languages traditionally support one indexing type: the natural numbers.
Here, we allow arbitrary ordered types.
Each {level} of a stream is annotated with a label.
We refer to a (label, type) pair as an \emph{attribute} or \emph{level}, and the sequence of attributes for a stream is its \emph{shape}.

\no{
\paragraph{Technical note:}
The boolean parameter
indicates whether \code{seek} should advance up to ($\ge i$) or past ($> i$) the given index.
In particular, we expect the following identity to hold between \code{seek} and the previous definition of \code{next}:
\code{next q = seek (index q) (ready q)}.
That is, if the state $q$ is ready, \code{seek} should make progress towards a state with index strictly greater than index $q$, whereas if $q$ is not ready, it must not advance past a state with index = \code{index q}, if it exists.
}

\end{document}